\documentclass[aps,pra,twocolumn,showpacs,floatfix,superscriptaddress,reprint,nobalancelastpage]{revtex4-1}
\usepackage[pdftex]{color}
\usepackage{hyperref}
\usepackage[english]{babel}
\usepackage{amsfonts}
\usepackage{amsmath}
\usepackage{graphicx}
\usepackage[caption=false]{subfig}

\usepackage{soul}
\usepackage{color}
\usepackage{ulem}

\newcommand{\Ket}[1]{\left|#1\right>}
\newcommand{\Bra}[1]{\left<#1\right|}
\newcommand{\BraKet}[2]{\left<#1|#2\right>}

\begin{document}

\title{Optimal control of many-body quantum dynamics: chaos and complexity}

\author{P. M. Poggi}
\email{ppoggi@df.uba.ar}
\affiliation{Departamento de F\'{\i}sica Juan Jose Giambiagi 
 and IFIBA CONICET-UBA,
 Facultad de Ciencias Exactas y Naturales, Ciudad Universitaria, 
 Pabell\'on 1, 1428 Buenos Aires, Argentina}
\author{D. A. Wisniacki}
\affiliation{Departamento de F\'{\i}sica Juan Jose Giambiagi 
 and IFIBA CONICET-UBA,
 Facultad de Ciencias Exactas y Naturales, Ciudad Universitaria, 
 Pabell\'on 1, 1428 Buenos Aires, Argentina}

\date{\today}

\begin{abstract}

Achieving full control of the time-evolution of a many-body quantum system is currently a major goal in physics. In this work we investigate the different ways in which the controllability of a quantum system can be influenced by its complexity, or even its chaotic properties. By using optimal control theory, we are able to derive the control fields necessary to drive various physical processes in a spin chain. Then, we study the spectral properties of such fields and how they relate to different aspects of the system complexity. We find that the spectral bandwidth of the fields is, quite generally, independent of the system dimension. Conversely, the spectral complexity of such fields does increase with the number of particles. Nevertheless, we find that the regular o chaotic nature of the system does not affect signficantly its controllability. 
\end{abstract}

\maketitle

\section{Introduction}\label{sec:intro}

The precise manipulation of nano and sub-nanoscale physical systems lies at the heart of the ongoing quantum revolution, by which new communication and information technologies are expected to emerge \cite{bib:natphot2009,bib:nature2010}. In this context, an amazing progress has been made in the study of non-equilibrium dynamics of many-body quantum systems, both  theoretically and experimentally \cite{bib:rev_mbody2008,bib:rev_neq2011}. A wide range of different phenomena has been closely studied in recent years, such as many-body localization \cite{bib:mbl2010,bib:mbl2015}, relaxation \cite{bib:rigol2007,bib:eisert2008,bib:wisn2015}, thermalization \cite{bib:rigol2008,bib:santos2011,bib:fazio2011}, quantum phase transitions \cite{bib:qpt2011}, among others.\\

Understanding the dynamics of such complex quantum systems is the first step towards the ultimate goal: the ability to engineer its complete time-evolution using a small number of properly tailored control fields. To tackle this problem, optimal control theory (OCT) \cite{bib:tannor1993,bib:rabitz1998} emerges as the natural tool. Routinely used in various branches of science \cite{bib:krotov1996}, optimization techniques allows to derive the required shape for a control field $\epsilon(t)$ that optimizes a particular dynamical process for a quantum system described by a Hamiltonian $H(\epsilon)$. For example, a typical goal in quantum control is to connect a given initial $\Ket{\psi_0}$ and target states $\Ket{\psi_f}$ in some evolution time $T$. In recent years, optimal control has been applied with great success in systems of increasing complexity, with applications including state control of many-boson dynamics \cite{bib:sherson2013,bib:calarco2015}, the crossing of quantum-phase transitions \cite{bib:doria2011}, generation of many-body entangled states \cite{bib:mintert2010,bib:caneva2012} and optimization of quantum thermodynamic cycles \cite{bib:montangero2016}. A lot of attention has also been devoted to investigate the fundamental limitations of OCT, most of all in connection with the study of the so-called quantum speed limit \cite{bib:caneva2009,bib:murphy2010,bib:hegerfeldt2013,bib:nos_qsl2013,bib:nos_qsl2015}. In a recent work, OCT has even been used in a citizen-science scenario allowing to investigate the power of gamification techniques in solving quantum control problems \cite{bib:sherson2016}.\\

In this work, we investigate the connection between the complexity of a quantum system and its controllability. To this end, we study optimal control protocols on a spin-1/2 chain with short-range interactions, both in the few- and many-body regimes. By using this model, we are able to tune the physical complexity of the system in two different ways: (a) by adding excitations to the chain, we can increase the system space dimension; (b) by tuning the interparticle coupling, we can drive the system through a transition from a regular energy spectrum to a chaotic one. We perform an unconstrained optimization in order to obtain the control fields needed to drive various physical processes, and define two figures of merit based on the frequency spectrum of the fields: the spectral bandwidth, associated with the maximum frequency present in the field and the spectral inverse participation ratio (sIPR), related to the signal complexity. We find that the spectral bandwidth is strongly connected to the structure of the control Hamiltonian. In the common scenario where the control is applied locally on any site of the chain, we find that the bandwidth is independent of the state space dimension, for various processes. On the other hand, the complexity of the signal grows with the dimension, due to the increase of energy levels. Inspired by this, we asses the role of quantum chaos in the control of quantum many-body dynamics. We find that the transition between regular and chaotic energy spectrum does not affect the complexity of the control problem in a significant way.\\

We point out that previous works have studied the relationship between optimal control and the integrable or non-integrable nature of the quantum system under analysis \cite{bib:caneva2014,bib:lloyd2014}. There, a suitable measure for the control field complexity was defined, related with the number of frequencies  in the field, as allowed by the optimization procedure. Then, it was shown that the complexity required to achieve control scaled exponentially with the dimension of the manifold supporting the dynamics. Here, we focus on studying the complexity of the control fields regardless of the details of the optimization method itself. We do this by deliberately allowing many frequency components in the control fields, and then analysing which of those components are required to effectively drive the system. \\

This article is organized as follows. In Sec. \ref{sec:model} we present the model of a spin-1/2 chain and discuss its symmetries. Also, we propose a few control protocols and put forward the main elements of QOC theory.  In Sec. \ref{sec:result} we define two measures of complexity for the optimal control fields, and present a systematic study on how they vary with both the system state dimension and chaos parameter. In. Sec. \ref{sec:discuss} we analyse in detail the reasons why the chaotic regime does not affect the controllability of the system. Finally, Sec. \ref{sec:conclu} contains some concluding remarks. 

\section{Model and methods} \label{sec:model}

\subsection{Spin chain model}

Let us consider a one-dimensional system of $L$ spin-1/2 particles that interact through nearest-neighbor (NN) and next-to-nearest-neighbor (NNN) homogenous couplings with open boundary conditions. The Hamiltonian for this models reads

\begin{eqnarray}
 H_{01} &=& H_0 + \Gamma H_1, \label{ec:h01}\\
 H_0 &=& \frac{J}{2}\sum_{i=1}^{L-1}\sigma_i^x\sigma_{i+1}^x+\sigma_i^y\sigma_{i+1}^y+\alpha_z\sigma_i^z\sigma_{i+1}^z,\\
 H_1 &=& \frac{J}{2}\sum_{i=1}^{L-2}\sigma_i^x\sigma_{i+2}^x+\sigma_i^y\sigma_{i+2}^y+\alpha_z\sigma_i^z\sigma_{i+2}^z,
\end{eqnarray}

\noindent where $\sigma_i^{x,y,z}$ are the Pauli matrices for the i-th particle and we have taken $\hbar=1$. The Hamiltonian $H_0$, which has only NN couplings, is the usual $XXZ$ Heisenberg model, which can be exactly solved via the Bethe ansatz \cite{bib:bethe}. The parameter $\Gamma$ measures the ratio between the NNN exchange and the NN couplings. This model has been extensively studied in the literature in many contexts \cite{bib:vedral2001,bib:fazio2002}, in particular when investigating quantum chaos \cite{bib:wisn2015,bib:santos2012}, i.e., the study of the quantum mechanical properties of systems which classical analogs display a chaotic behavior. Albeit lacking a semiclassical counterpart, this spin model displays a transition in its level spacing distribution as $\Gamma$ changes. For $\Gamma\lesssim0.5$, the energy spectrum is regular, and its level spacings follow a Poisson distribution, while for $\Gamma\gtrsim0.5$, the distribution follows Wigner-Dyson statistics, and the spectrum is deemed "chaotic" \cite{bib:stockman,bib:bohigas} (see the Appendix for more details).\\

While the full Hilbert space $\mathcal{H}$ of this model has a dimension of $2^L$, we can identify two constant of motions such that $\mathcal{H}$ is decomposed into smaller subspaces. First, the total magnetization the in $z$ direction $\sigma_z=\sum_{i=1}^L \sigma_i^z$  is conserved, which defines $L+1$ subspaces with fixed $\sigma_z$. Each subspace can then be characterized by this quantum number, which can be interpreted as the number of sites with spins pointing ``up'' or ``excitations'' in the chain. The dimension of the subspace with $K$ excitations is given by
\begin{equation}
  D_K=\frac{L!}{K!(L-K)!}.
\end{equation}

We consider also conservation of parity. The parity operator $\Pi$ acts as permutation between mirrored sites of the chain, and commutes with the Hamiltonian $H_{01}$ for all values of $J$ and $\Gamma$. As a consequence, each of the above mentioned subspaces break up into two (positive and negative) parity subspaces, each of which of dimension $D_{K,\Pi}\simeq D_K/2$ \cite{bib:santos2012}. Lastly, we avoid conservation of $\sigma_z^2$ by chosing $\alpha_z=0.5$, and choose an odd value of the chain length $L=15$.

\subsection{Control protocols}

We now describe an scenario to perform control operations on the spin chain. First, we define the control Hamiltonian by means of which we intend to steer the ``free'' chain Hamiltonian $H_{01}$. Several different proposals have been studied in the literature. For example, in Ref. \cite{bib:bose2007}, the author proposed using a global parabolic magnetic potential to control the transfer of excitations from one end of the chain to the other. Later, the same configuration was used to study the optimal evolution time for such processes \cite{bib:caneva2009,bib:murphy2010}. Other control configurations have also been proposed in scenarios where only one \cite{bib:spin1} or two \cite{bib:spin2,bib:spin2b,bib:spin2c} sites are locally addressed by external fields.  Here, we will consider time-dependent magnetic fields in the $z$ direction which are locally applied at each site of the chain. In order to comply with the system symmetries, we will consider the situation where the first and last spin of the chain are affected by the same field $\epsilon(t)$, whereas all the other spins do not interact with any external field. Consequently, the full Hamiltonian can be written as

\begin{equation}
  H(t) = H_{01} + \epsilon(t)H_c,\ \mathrm{where}\ H_c=\frac{J}{2}\left(\sigma_1^z+\sigma_L^z\right).
  \label{ec:hcontrol}
\end{equation} 

We point out that $H(t)$ preserves the same symmetries as $H_{01}$ for any choice of $\epsilon(t)$. In addition to this, we checked that for any fixed value of the field, $H$ still shows a transition between a regular and a chaotic spectrum for $\Gamma\simeq0.5$. For more details about this issue, see the Appendix.\\

The next step is to define the control processes we aim to perform. We will consider two different protocols (A and B) in order to obtain general results about the systems controllability. In both cases, we define initial and target states which we denote $\Ket{\psi_0^\alpha}$ and $\Ket{\psi_f^\alpha}$, where $\alpha=A,B$. These states are deliberately defined to allow the system evolve within a particular subspace with fixed (positive) parity and number of excitations $K$ of the complete Hilbert space, as discussed in the previous section. First, process ``A''  involves the system initially prepared in a state with all excitations in the middle sites of the chain (in this scheme, the central site has no excitations if K is even).  We then intend to drive this configuration into a coherent superposition as defined by

\begin{eqnarray}
  \Ket{\psi_0^A}&=&\Ket{\downarrow\ldots\downarrow\uparrow\uparrow\uparrow\downarrow\ldots\downarrow}\\
  \Ket{\psi_f^A}&=&\frac{1}{\sqrt{2}}\left(\Ket{\uparrow\uparrow\uparrow\downarrow\ldots\downarrow}+\Ket{\downarrow\ldots\downarrow\uparrow\uparrow\uparrow}\right).
\end{eqnarray}

Process A then represents and orderered control process in which entanglement is generated between both ends of the chain. On the other hand, we define a disordered process B, where the system starts from the ground state of $H_0$ and its steered into a random superposition of excited states (with positive parity). 

\begin{eqnarray}
  \Ket{\psi_0^B}&=&\Ket{\mathrm{g.s.}_0}\\
  \Ket{\psi_f^B}&=&\sum_{n=1}^{D_{K,+}-1}a_n\Ket{n_0}
\end{eqnarray}

\noindent where $\left\{\Ket{n_0}\right\}$, $n=0,\ldots,D_{K,+}$ are the positive eigenvectors of $H_0$ in the subspace of $K$ excitations and $\Ket{\mathrm{g.s.}_0}\equiv\Ket{0_0}$. The coefficientes $\left\{a_n\right\}$ are a set of random complex numbers so that $\sum_i |a_n|^2=1$.\\

\subsection{Optimal control methods}

In order to obtain the control fields $\epsilon(t)$ which drive the system for both processes, we use optimal control theory. Here we briefly sketch the Krotov optimization algorithm, as described in many previous works (see for example Refs. \cite{bib:krotov1996,bib:sklarz2002,bib:montangero2007,bib:werschnik2007,bib:nos_oc2015}). This procedure takes as an input a fixed evolution time $T$, an initial guess for the control field $\epsilon^{(0)}(t)$, and both the initial and final states $\Ket{\psi_0^\alpha}$ and $\Ket{\psi_f^\alpha}$, where $\alpha=A,B$. The procedure starts by evolving $\Ket{\psi_0^\alpha}$ according to the Hamiltonian $H(\epsilon^{(0)}(t))$, from $t=0$ to $t=T$. The final state $\Ket{\psi(T)}$ is then projected to $\Ket{\psi_f^\alpha}$ in order to obtain an auxiliary state $\Ket{\chi(T)}=\BraKet{\psi_f^\alpha}{\psi(T)}\Ket{\psi_f^\alpha}$. This state is finally evolved backwards with the same Hamiltonian, from $t=T$ to $t=0$. The process is then repeated, but the control field is updated following the recipe

\begin{equation}
  \epsilon^{(k+1)}(t)\rightarrow\epsilon^{(k)}(t) + \frac{1}{\lambda(t)}\mathrm{Im}\left\{\Bra{\chi(t)}\frac{\partial H}{\partial \epsilon}\Ket{\psi(t)}\right\},
\end{equation}

\noindent where $\lambda(t)$ is an weight function. Note that, for the model considered here, the operator $\partial H/\partial \epsilon$ is fully defined by equation (\ref{ec:hcontrol}) and equals simply to $H_c$. The iterative procedure stops when a certain target fidelity $\mathcal{F}=|\BraKet{\psi(T)}{\psi_f}|^2$ has been achieved.\\

\begin{figure}[!t]
\includegraphics[width=\linewidth]{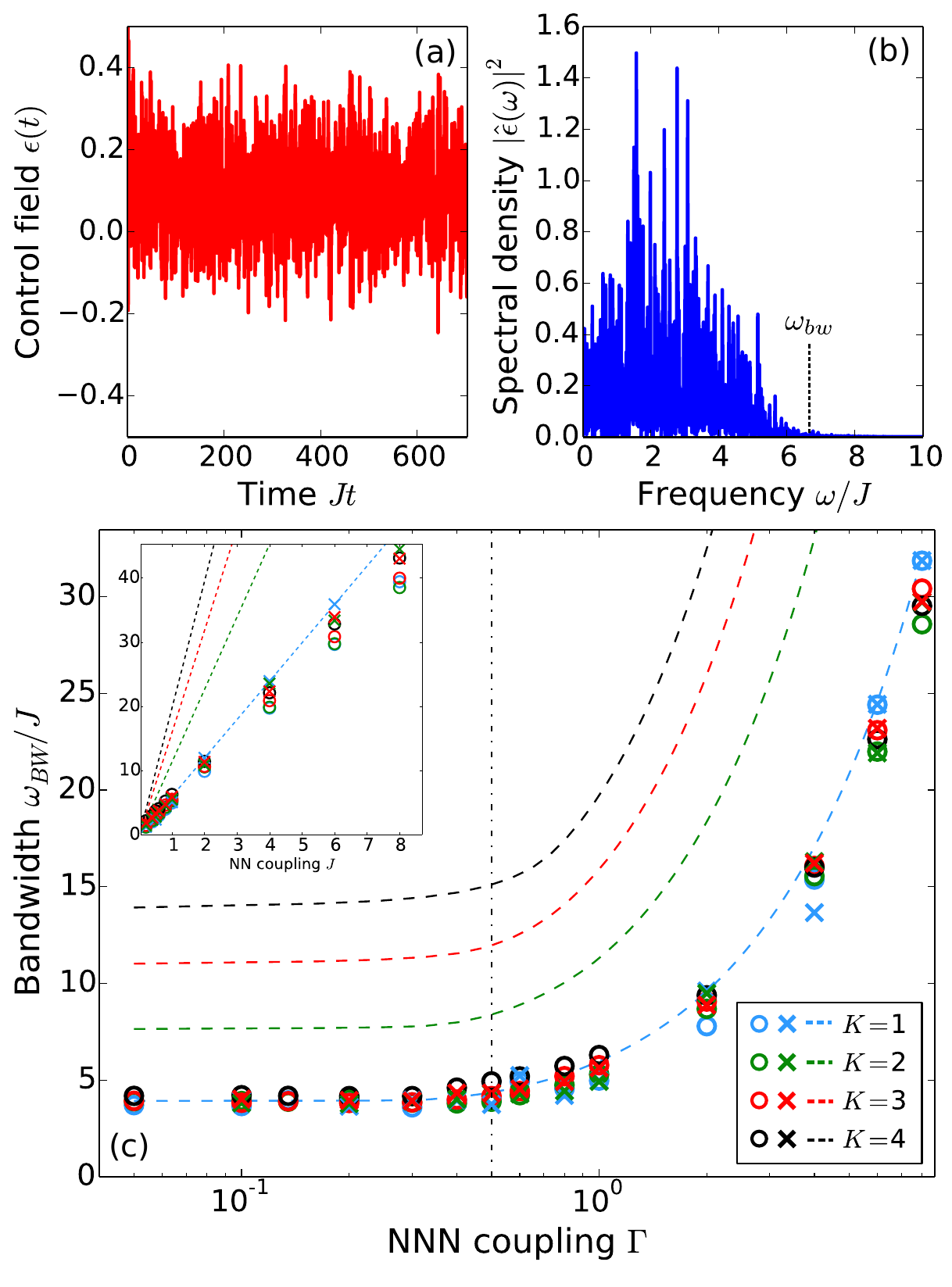}
\caption{\label{fig:fig1} (color online) (a) A typical optimized control field $\epsilon(t)$ obtained for process A and $\Gamma=1$ and (b) its Fourier transform. (c) Spectral bandwidth of the optimized control fields as a function of the adimensional NNN coupling $\Gamma$, for various values of $K$ (the number of excitations in the spin chain). Data shown is for both control processes A and B (see text for details). State space dimension ranges from 15 (K=1) to 1365 (K=4). The dashed curves show the energy spread of the free Hamiltonian $H_{01}$ (\ref{ec:h01}) for different values of $K$. The dash-dotted vertical line indicates the critical value of $\Gamma=0.5$ for which the regular-chaotic transition ocurrs in the energy spectrum of $H_{01}$. Inset: same as in the main figure, but plotted as a function of the NN coupling J, for a  fixed value of $\Gamma=1$.}
\end{figure}

As we intend to compare the optimal control fields obtained by this optimization procedure, we fix the input parameters of the optimization as follows. For the total evolution time $T$ we set $T=15\times T_L$ where $T_L=(L-1)\frac{\pi}{J}$ can be regarded as the typical evolution time required for transferring a single excitation from one end of the chain to the other \cite{bib:murphy2010}. We have checked that using this value we are operating well beyond the quantum speed limit \cite{bib:caneva2009}, and so that fidelities up to 0.99 or greater can be achieved, for both control processes and every value of the number of excitations $K$ and the NNN coupling $\Gamma$ considered. Also, we used a constant initial guess $\epsilon^{(0)}(t)=0.1$ in all cases. We have checked that the results we present in the next section hold for other choices of this function.


\section{Analysis of the optimal control fields} \label{sec:result}

In Fig. \ref{fig:fig1} (a) we show a typical example of the control field $\epsilon(t)$ obtained by the optimization procedure outlined in the previous section, together with its Fourier spectrum. The time signal shown can be seen to be complex and to have many spectral components up to certain frequency threshold. In order to characterize quantitively this features, we define two measures of complexity for the control fields: the frequency bandwidth and the spectral inverse participation ratio (sIPR). In this section we investigate these quantites.

\subsection{Frequency bandwidth}

\begin{figure}[!b]
\includegraphics[width=\linewidth]{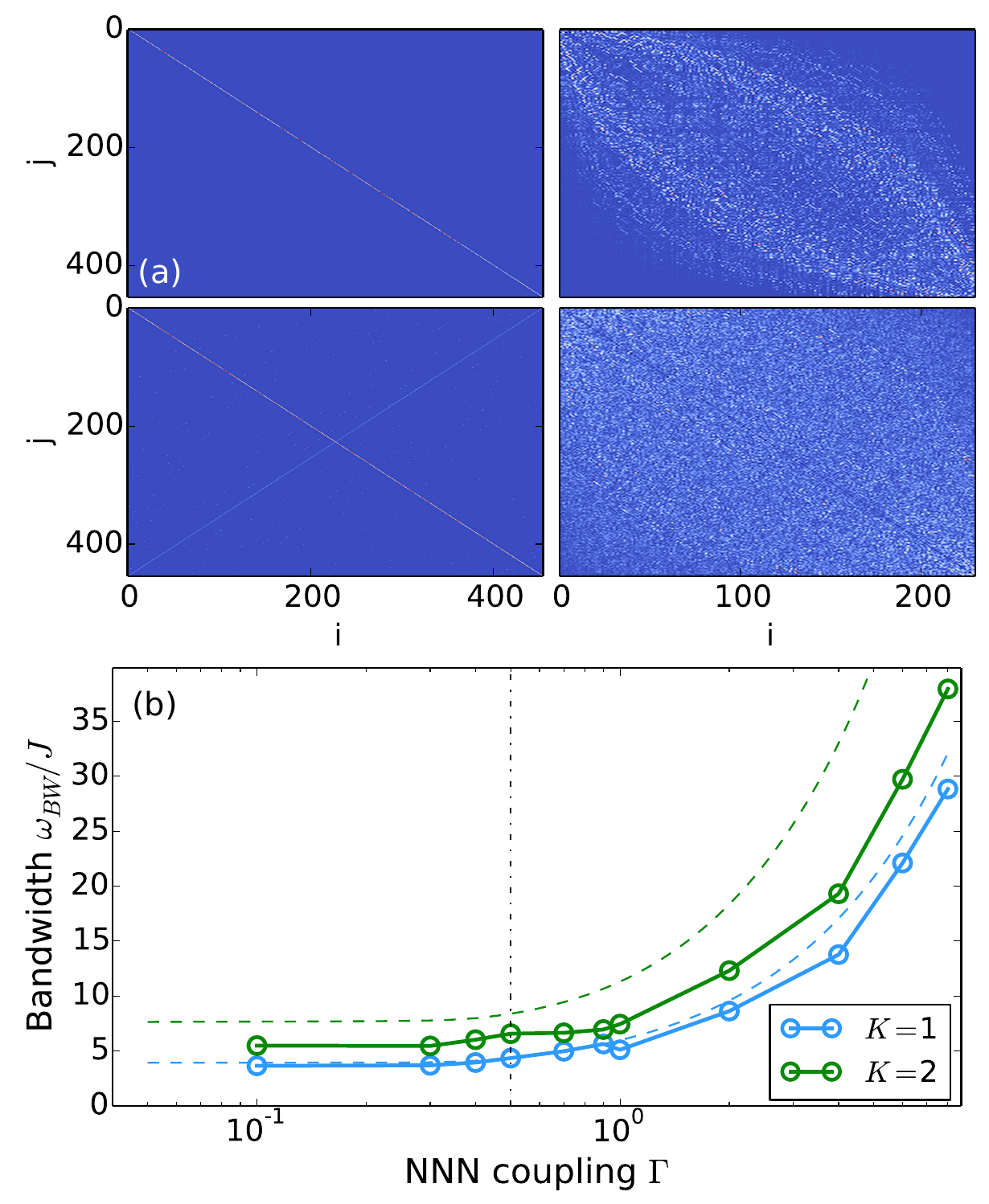}
\caption{\label{fig:fig2} (color online) (a) Absolute values of the matrix elements for the control Hamiltonian in the computational basis (left panels) and in the energy eigenbasis with positive parity (right panels). Top panels: local control with $H_c$ as in eq. \ref{ec:hcontrol}. Bottom panels: long-range control with $H'_c$ described in text. (b) Frequency bandwidth of the optimized control fields as a function of the adimensional NNN coupling $\Gamma$, for $K=1,\:2$ (the number of excitations in the spin chain). Data shown is for process A (see text for details), using the long range control Hamiltonian $H'_c$. }
\end{figure}

Given a time-varying signal $\epsilon(t)$ and its Fourier transform $\hat{\epsilon}(\omega)$, we first define its frequency bandwidth as the value $\omega_{bw}$ such that

\begin{equation}
  \int_0^{\omega_{bw}}d\omega\:|\hat{\epsilon}(\omega)|^2=1-\beta,
  \label{ec:bw}
\end{equation}

\noindent where $0<\beta<1$ and the frequency distribution is normalized such that $\int_0^{\infty}d\omega\:|\hat{\epsilon}(\omega)|^2=1$. By this definition, the frequency interval $[0,\omega_{bw}]$ concentrates the $[(1-\beta)\times100]\%$ of the power spectrum (here, we use $\beta=10^{-2}$). In other words, $\omega_{bw}$ is a measure of the maximum frequency present in $\epsilon(t)$. \\

In Fig. \ref{fig:fig1} (b) we show the frequency bandwidth $\omega_{bw}$ as a function of the NNN or chaos parameter $\Gamma$, for different number $K$ of excitations in the chain. Results obtained for both processes A and B are shown in the same plot. Remarkably, we find that all data roughly coincides in the same curve. This result indicates that the bandwidth is independent not only of the control processes considered, but also of the state space dimension. Note that, in each case, $\omega_{bw}$ is approximately constant for $\Gamma<0.5$ and then increases steadly for $\Gamma>0.5$. Although this behavior correlates with the onset of chaos in the system (as discussed in Sec. \ref{sec:intro}), we must first consider that increasing the interparticle coupling $\Gamma$ necessarily increases the energy of the chain. As previously discussed in the context of QOC \cite{bib:rabitz1998_2,bib:tesch2002}, we expect that the frequency distribution of the control fields presents peaks located at the resonances of the free Hamiltonian $H_{01}$. Following this criterion, the maximum frequency is bounded by the energy spread $\Delta E$ of $H_{01}$, defined as

\begin{equation}
  \Delta E = E_{max}-E_{0},
\end{equation}

\noindent where $E_{max}$ and $E_0$ are the maximum and minimum (ground state) energies of the Hamiltonian. Note that $\Delta E$ is a function of the interparticle interaction parameters $J$ and $\Gamma$ and of the number of excitations $K$. We show such functions as dashed lines in Fig. \ref{fig:fig1} (c). It is clear that the dependence of the bandwidth with $\Gamma$ closely resembles the energy spread with $K=1$. The same observation can be drawn by studying both quantities as a function the NN coupling $J$ (for fixed $\Gamma$). There, the behaviour is obviously linear, as shown in the inset of Fig. 1 (c).\\

Note, however, that the striking independence of the bandwith with the state dimension cannot be explained by its relation with the energy spread of the free Hamiltonian, since $\omega_{bw}$ is in every case significantly smaller than $\Delta E$ for $K>1$. In order to gain a deeper insight about this result, we turn to investigate the role of the control Hamiltonian $H_c$, defined in eq. (\ref{ec:hcontrol}). We first study the structure of the matrix $H_c$ written in the basis of (positive) eigenvectors of the free chain Hamiltonian $H_{01}$. In the top row of Fig. \ref{fig:fig2} (a) we plot the absolute value of such matrix elements for fixed values of $\Gamma$ and $K$. From this plots we can see that $H_c$ does not connect eigenstates which distant energies: for example, the ground state is not connected with excited states beyond the middle of the spectrum. This explains the absence of  such high transition frequencies in the spectrum of the control fields. In order to provide numerical proof about this feature, we studied the implementation of one the control processes with a different choice of control operator $H'_c$ which presents a higher connectivity \cite{bib:santos2012} between distant states in the spectrum. Such Hamiltonian matrix is shown in the bottom-right panel of Fig. \ref{fig:fig2} (a). Results for the new optimized fields are shown in Fig. \ref{fig:fig2} (b), where we show the frequency bandwidth as a function of $\Gamma$ for this case. It can be seen that $\omega_{bw}$ is greater for $K=2$ than for $K=1$, for all values of $\Gamma$ considered. We point out that, by looking at the representation of $H'_c$ in the computational basis (bottom left panel), we can see that this alternative control procedure would involve tuning a complex combination of multi-spin interactions, in clear contrast with the simple structure of $H_c$. \\

The results shown so far allows us to assert that the control bandwidth, which measures the range of frequencies present in the fields is determined exclusively by the energy spread of the free Hamiltonian and the structure of the control Hamiltonian. This gives us a measure of the physical complexity of the control field which turns to be independent of the number of particles in the system. We point out here that we are not interested in analyzing the complexity of the optimization itself, as has been done in previous works which have obtained interesting results \cite{bib:caneva2014,bib:lloyd2014}. We work our way around this issue by fixing the time step of our numerical implementation to very small values, $J\Delta t = 10^{-2}$. This determines that the maximum allowed frequency in the fields is en every case at least on order of magnitud higher than the actual physical frequencies found by Fourier analysis in the control fields.

\subsection{Spectral localization}

\begin{figure}[!b]
\includegraphics[width=\linewidth]{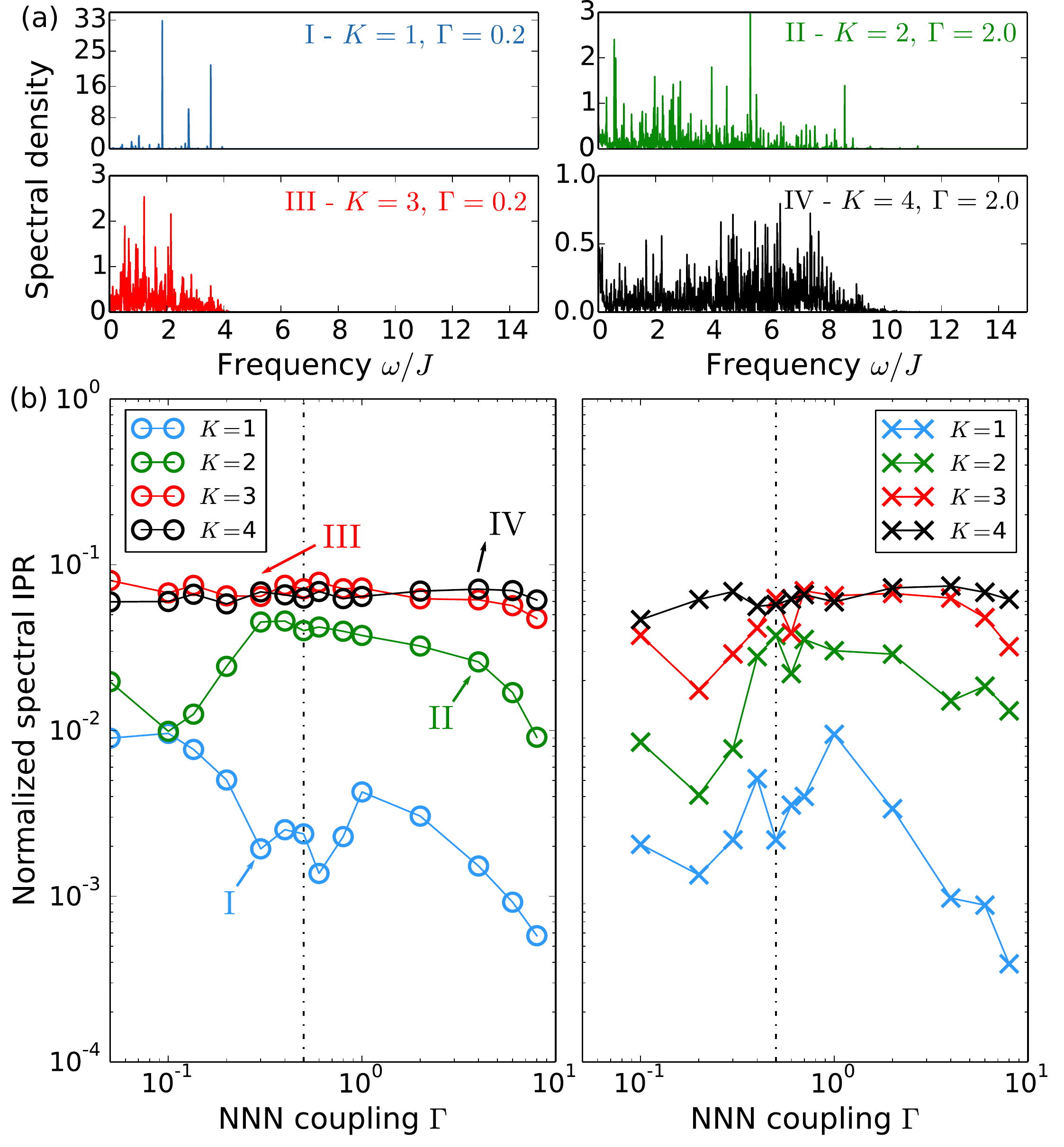}
\caption{\label{fig:fig3} (color online) (a) Various spectral densities of the optimal control fields found for process A. Values for the excitation number $K$ and the adimensional next to nearest-neighbour coupling $\Gamma$ are shown for each case I-IV. (b) Normalized spectral inverse participation ratio (sIPRn), which measures the control field complexity, as a function of $\Gamma$, for various values of $K$ (the number of excitations in the spin chain). Data shown is for process A (left) and B (right). The points indicated by labels I-IV correspond to the spectrums shown in (a). State space dimension ranges from 15 ($K=1$) to 1365 ($K=4$). The dash-dotted vertical line shows the critical value of $\Gamma$ for the regular-chaotic transition in the energy spectrum.}
\end{figure}

We now turn our attention to another measure of the control field complexity. In this case, we to study how the number of frequencies which appear in the signal spectrum varies as the systems complexity is increased. For this purpose, we define the following quantity

\begin{equation}
  \mathrm{sIPR} = \left(\int_0^\infty d\omega\:|\hat{\epsilon}(\omega)|^4\right)^{-1},
\end{equation}

\noindent which we call ``spectral inverse participation ratio'' (sIPR) as it is inspired in the commonly known IPR \cite{bib:wisn2015,bib:ipr1,bib:ipr2}. The sIPR quantifies the localization in the Fourier transform of a time signal, and thus allows us to asses how complex the control field is inside its bandwidth. Note that localized frequency spectrums give sIPR$\rightarrow0$, and complex signals with delocalized spectrum tend to higher sIPR. As an example, take a completely random signal with frequency components up to $\omega_{bw}$. We expect such a signal to have a flat Fourier transform $\hat{\epsilon}(\omega)=1/\omega_{bw}$ for $0<\omega<\omega_{bw}$ and $\hat{\epsilon}(\omega)=0$ for $\omega>\omega_{bw}$. Calculating the sIPR in that case is straightforward and gives $\omega_{bw}$. We point out that here we intend to quantify the optimal control field complexity regardless of the frequency distribution width. For this purpose, we evaluate the normalized sIPR

\begin{equation}
  \mathrm{sIPRn} = \frac{\mathrm{sIPR}}{\omega_{bw}}.
\end{equation}

Following the discussion on the previous paragraph, we expect sIPRn to range between 0 and 1, and we can interpret it as a measure of resamblence between the signal under study and a completely random time field.\\

In Fig. \ref{fig:fig3} we plot the normalized spectral IPR as a function of the NNN exchange $\Gamma$ for different values of the number $K$ of excitations in the spin chain, and for both control processes A and B. We show also some examples of the frequency spectrums we obtained, and it can be corroborated that sIPRn effectively measures how localized the spectrum is in Fourier space. More generally, it can be seen from the figure that sIPRn takes small values for $K=1$ and then grows with $K$, and thus with state space dimension of the system. This is in sharp contrast with the behaviour of the frequency bandwidth $\omega_{bw}$, which was found to be independent of $K$. We point out that this behaviour is common to both control processes. It is interesting to note that the high-dimensional cases ($K=3,4$) roughly converge to the same value of sIPRn, indicating that there maybe an upper bound for this measure which is below its maximal theoretical sIPRn$=1$, which is achieved when the frequency spectrum is flat. Physically, the existence of an upper bound $<1$ means that optimal control fields can always be distinguished from completely random, white noise-type fields. We leave this issue for future investigation.\\

Despite the dependance of normalized spectral IPR with the space dimension, it can be seen also that this indicator does not exhibit any clear trend with the NNN parameter $\Gamma$. We observe that, for small values of $K$, this parameter shows large fluctuations which tend to attenuate when for larger space dimensions. We recall that, for $K\geq 3$, the system exhibits a clear transition from a regular energy spectrum to a more complex (chaotic) one at $\Gamma=0.5$. As can be seen from Fig. \ref{fig:fig3}, there is no evidence of such leap in complexity in our numerical study. In this way, we can assert that the optimal fields required to control the dynamics of regular or chaotic Hamiltonians display a similar spectral complexity.\\

As a final remark, we point out that it would not be correct to claim that the spectral properties analized in this section are completely independent of the choice of initial and final state. This can be easily seen by considering a processes where we intend to connect the ground state of the free Hamiltonian $H_{01}$ and one of its excited states $\Ket{n(\Gamma)}$. If the control Hamiltonian $H_c$ connects both states, we expect that the bandwidth of the control field will be given by the energy difference between both levels $E_n(\Gamma)-E_0(\Gamma)$, which can be signficantly lower than the obtained $\omega_{bw}$ for processes A and B if $\Ket{n}$ lies in the low-energy region. Nevertheless, our results do apply to general linear combinations of energy eigenstates, which is the more common scenario. 

\section{Discussion} \label{sec:discuss}

\begin{figure}[b]
\includegraphics[width=\linewidth]{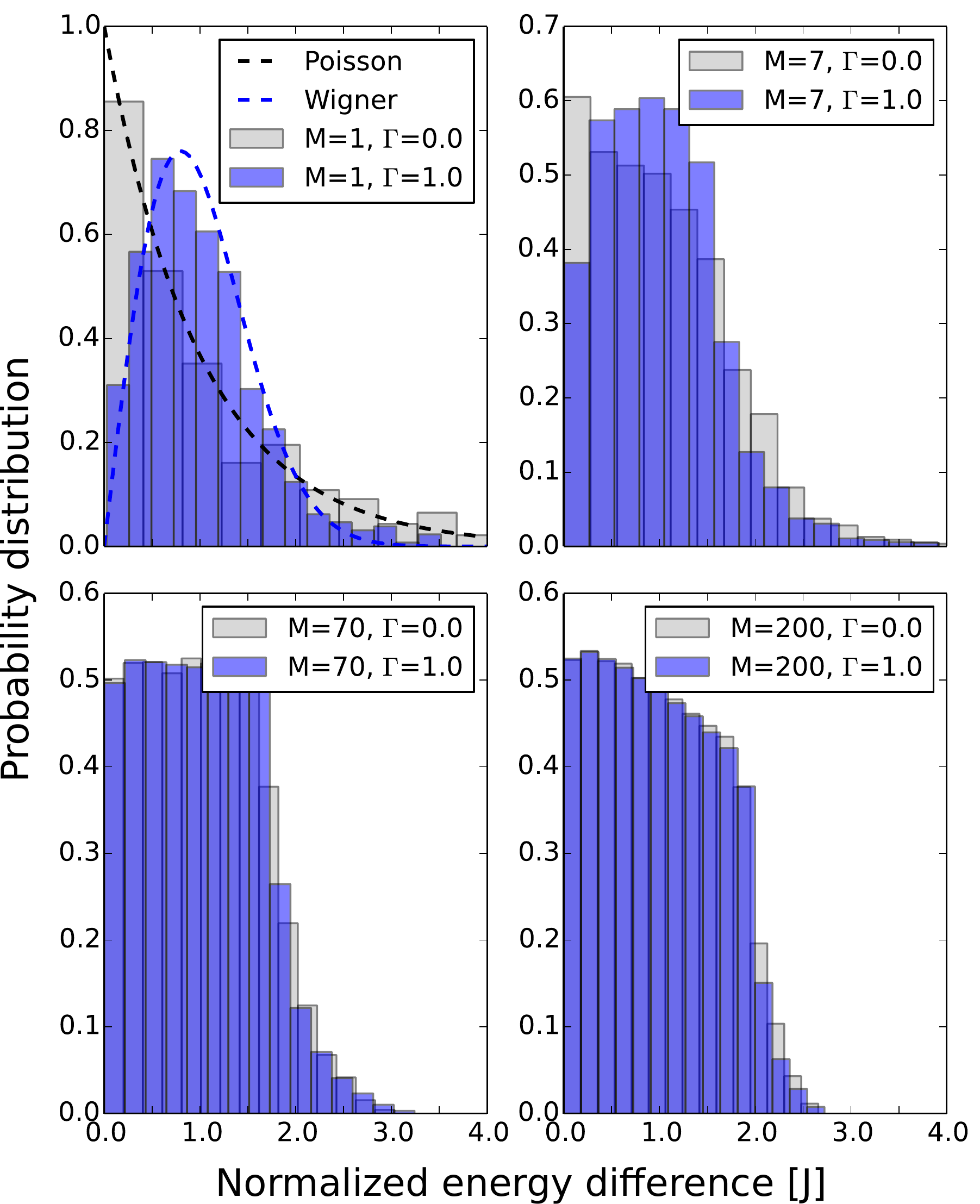}
\caption{\label{fig:fig4} (color online) Distribution of energy differences $\delta E_{n,m}$ defined in eq. (\ref{ec:levelspm}) for $\Gamma=0$ (regular) and $\Gamma=1$ (chaotic), where $\Gamma$ is the NNN coupling parameter. All cases are normalized such that their mean value is 1 (in units of $J$). In the top left plot, we show the $M=1$ case corresponding to the standard level spacing distribution. Poisson and Wigner-Dyson distributions are shown on top of the histograms. In all cases, we consider the positive subspace with $K=4$ excitations in the chain ($D_{4,+}\simeq 700$). }
\end{figure}

We will now look more closely at the connection between the spectral features of the optimal control fields and the structural properties of the system spectrum. We have already pointed out in Sec. \ref{sec:model} that the free chain Hamiltonian $H_{01}$ \ref{ec:h01} shows a transition in its level spacing distribution $\{\delta E_n\}$ as the NNN coupling parameter $\Gamma$ changes, where
\begin{equation}
  \delta E_n = E_{n+1}-E_n,
  \label{ec:levelsp1}
\end{equation}
\noindent and $E_n$ is the $n$th ordered eigenvalue of $H_{01}$. If the space dimension is high enough ($K\geq 3$), the level spacings statistics show a Poisson distribution for $\Gamma\lesssim0.5$, and a Wigner-Dyson distribution $\Gamma\gtrsim0.5$ (see the Appendix for more details). We have also discussed in Sec. \ref{sec:result} that we observed a connection between the frequency components present in the optimized control field and the resonances of the free Hamiltonian $H_{01}$. Thus, an interesting point arises: if the energy spectrum of $H_{01}$ changes its structure with $\Gamma$, why is there no evidence about those changes in the frequency distribution of the optimal control fields?\\

The key point here is to note that the resonances of $H_{01}$, which feed the frequency distribution of the control field, are not only formed by the difference of two consecutive energies $\delta E_n$ (\ref{ec:levelsp1}). If connected by the control Hamiltonian, every energy difference present in the spectrum is also a suitable candidate for appearing the control field frequency spectrum. Following this discussion, we studied the distribution of the energy differences defined as

\begin{equation}
  \delta E_{n,m} = E_{n+m}-E_n\ \mathrm{with}\ 0<m\leq M,
  \label{ec:levelspm}
\end{equation}

\noindent such that $\delta E_{n,1}\equiv \delta E_n$. Note that, for every $n$, the value of $M$ indicates how many levels above $E_n$ are considered, and is thus bounded by the space dimension $D_{K,+}$. In Fig. \ref{fig:fig4} we show the distributions of normalized energy differences for different values of $M$, using $\Gamma=0$ and $\Gamma=1$. There, it can be seen that both distributions show the expected Poisson and Wigner-Dyson shapes when $M=1$ (as discussed in the previous paragraph), but start to converge to a common form when $M$ grows. As an example, for $K=4$, we have that dimension of the positive subspace is $D_{4,+}\simeq700$, and already taking $M\simeq D_{4,+}/10$ already gives near perfectly matching distributions for both values of the chaos parameter $\Gamma$. This analysis indicates that, while level spacing distributions are quite different for regular and chaotic spectra, the overall energy difference distributions converge to a same shape. This interesting behavior determines that the frequency spectrum of the optimal field which control both type of systems have the same complexity.\\

\section{Final remarks} \label{sec:conclu}

In this work we studied control processes in a chain of spin-1/2 particles and investigated how the complexity of the physical system relates to the complexity of the control field. We studied a Heisenberg chain model, which allowed us to consider separately different space dimensions (ranging from $\sim10$ to $\sim700$ states) by adding excitations to the system. By allowing next-to-nearest neighbour interactions, we were also able to parametrically tune the system from regular to chaotic. We find the time-dependent control fields required to drive different processes using optimal control theory and defined two measures of complexity based on the Fourier spectrum of those fields. By doing so, we could identify which aspects of the systems complexity affect the control fields. For instance, we found that the spectral bandwidth, which measures the maximimum frequency present in the field, is quite generally independent of the system space dimension. However, we showed that exceptions to this rule ocurr if we choose highly non-local control fields. Also, we investigated how many frequencies present inside the signal bandwidth, by defining a measure of localization: the spectral inverse participation ratio (sIPR). We found that this measure of field complexity does increase when excitations are added to system. Finally, we assesed the role of quantum chaos in the control of the system by studying the fields as a function of the chaos parameter $\Gamma$. We found no evidence of the regular - chaotic transition in the field spectral measures, allowing us to assert that the fields required to control chaotic and integrable systems display the same complexity. Concerning the role of quantum chaos in the dynamics of many-body systems, it is interesting to point out that a previous work \cite{bib:wisn2015} studied relaxation processes in such systems. Although working in an opposite scenario to coherent control, the authors also found no trace of the chaoticity of the system in the relaxation dynamics. In our case, we present further evidence about the irrelevance of quantum chaos in the coherent dynamics of many-body system.  \\

\appendix*
\section{Regular to chaotic transition in the spin chain}

\begin{figure}[h]
\includegraphics[width=\linewidth]{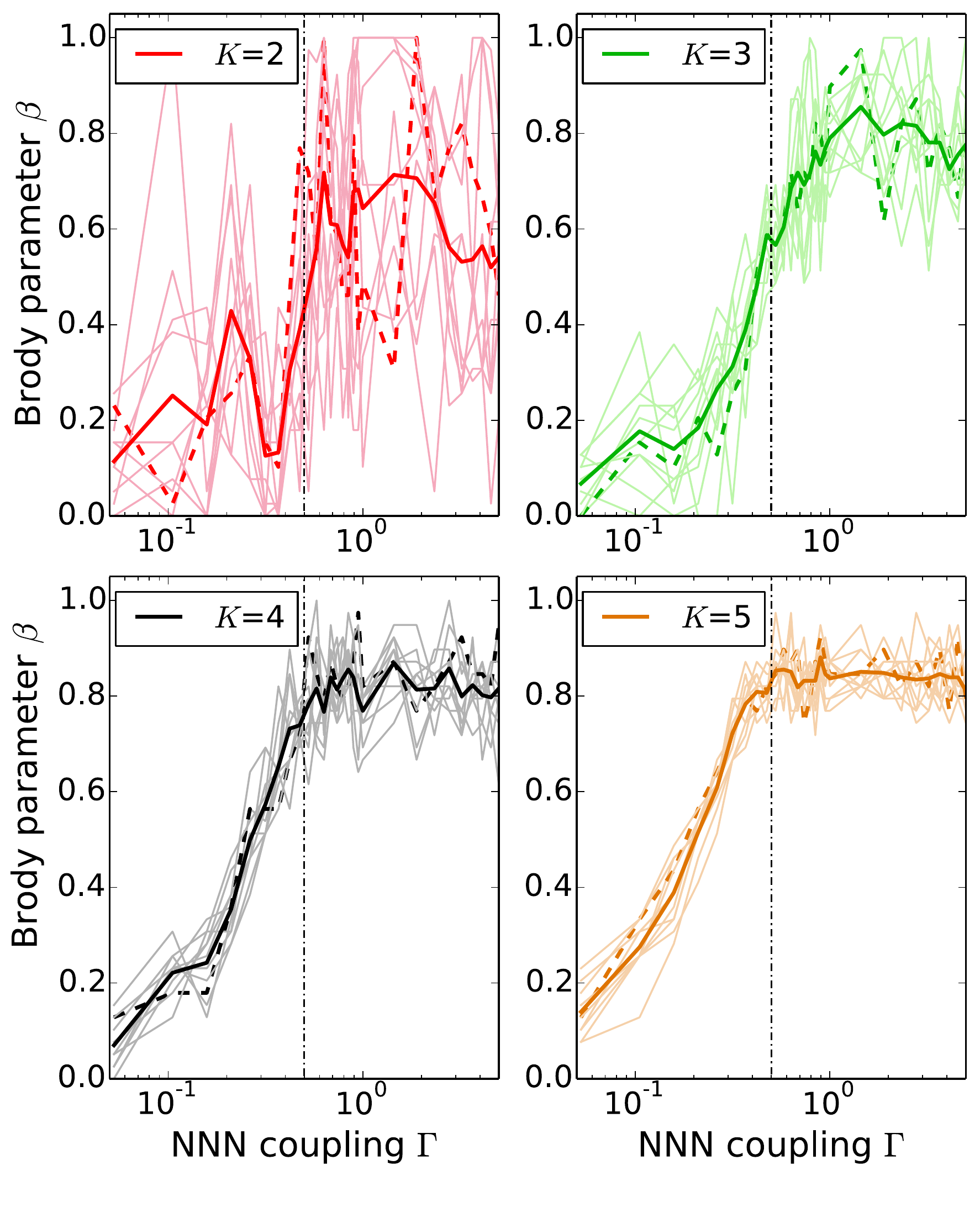}
\caption{\label{fig:fig5} (color online) Brody parameter $\beta$ as a function of the NNN coupling $\Gamma$ for the full spin chain Hamiltonian (\ref{ec:hcontrol}). The four plots correspond to various values of the number $K$ of excitations in the chain. In each case, curves in light colors correspond to the results obtained by fixing the adimensional control parameter $\epsilon$ to values in the range $\left[-3,3\right]$. The special case $\epsilon=0$ (where $H=H_{01}$) is shown with thick, dashed lines. Thick full lines denote the mean value of all curves. The dash-dotted vertical line shows the critical value of $\Gamma$ for the regular-chaotic transition in the energy spectrum }
\end{figure}

Analysing the level spacing distribution is the most commonly used method to identify whether a system shows integrability or quantum chaos. Integrable quantum systems can present crossings in its energy levels. Their spectrums are called regular, and follow a Poisson distribution, 

\begin{equation}
  P_P(s)=\mathrm{exp}(-s).
\end{equation}

In chaotic systems, on the other hand, level crossings are avoided. It turns out that their level spacing statistics can be fully predicted by random matrix theory \cite{bib:stockman,bib:bohigas}, leading to a Wigner-Dyson distribution

\begin{equation}
  P_{WD}(s)=\frac{\pi}{2}s\:\mathrm{exp}\left(-\frac{\pi}{4}s^2\right)
\end{equation}

An elegant way to quantify the level of chaoticity in a quantum system is by employing the Brody parameter $\beta$. This number is obtained by fitting the actual level spacing distribution $P(s)$ with the Brody distribution \cite{bib:brody},

\begin{equation}
  P_B(s) = (\beta+1)bs^\beta \mathrm{exp}(-bs^{\beta+1}),\ b=\left[\Gamma\left(\frac{\beta+2}{\beta+1}\right)\right]^{\beta+1}.
\end{equation} 

Note that $\Gamma(x)$ refers here to the usual gamma function. The function $P_B(s)$ tends to a Poisson distribution $P_P(s)$ for $\beta\rightarrow0$, while for $\beta\rightarrow1$ resembles the Wigner-Dyson distribution $P_{WD}(s)$.\\

In order to analyze the chaotic properties of our spin model, we calculated the Brody parameter for the energy spectrum of the full system Hamiltonian $H(\epsilon)=H_{01}+\epsilon H_c$. We recall that $H_{01}$ corresponds to the free chain term (\ref{ec:h01}) and $H_c$ to the control operator (\ref{ec:hcontrol}). Results for the parameter $\beta$ as a function of the NNN coupling $\Gamma$ are shown in Fig. \ref{fig:fig5} for several (fixed) values of the control parameter $\epsilon$. There, it can be seen that there is a clear crossover from integrability to quantum chaos in our model for $\Gamma\simeq 0.5$, for any value of $\epsilon$. It is important to point out that this conclusion is only valid for $K\geq 3$, where the amount of energy levels is considerable and thus a statistical analysis of the energy levels can be performed. 

\begin{acknowledgments}

We acknowledge support from CONICET, UBACyT, and ANPCyT (Argentina). We would also like to acknowledge Augusto Roncaglia for useful discussions.

\end{acknowledgments}


\begin{thebibliography}{60}
\small{

\bibitem{bib:natphot2009} J.L. O’Brien, A. Furusawa, and J. Vuckovic, Nature Photon. \textbf{3}, 687–695 (2009).

\bibitem{bib:nature2010}  T.D. Ladd, F. Jelezko, R. Laflamme, Y. Nakamura, C. Monroe, and J.L. O’Brien, Nature \textbf{464}, 45 (2010).

\bibitem{bib:rev_mbody2008} I. Bloch, J. Dalibard, and W. Zwerger, Rev. Mod. Phys. \textbf{80}, 885  (2008).

\bibitem{bib:rev_neq2011} A. Polkovnikov, K. Sengupta, A. Silva, and M. Vengalattore, Rev. Mod. Phys. \textbf{83}, 863 (2011).

\bibitem{bib:mbl2010} A. Pal and D.A. Huse, Phys. Rev. B \textbf{82}, 174411 (2010).

\bibitem{bib:mbl2015} P. Bordia, H.P. L\"uschen, S.S. Hodgman, M. Schreiber, I. Bloch, and U. Schneider, arXiv:1509.00478 (2016).

\bibitem{bib:rigol2007} M. Rigol, V. Dunjko, V. Yurovsky, and M. Olshanii, Phys. Rev. Lett. \textbf{98}, 050405 (2007).

\bibitem{bib:eisert2008} M. Cramer, A. Flesch, I.P. McCulloch, U. Schollwöck, and J. Eisert, Physical Rev. Lett. \textbf{101}, 063001 (2008).

\bibitem{bib:wisn2015} I. García-Mata, A.J. Roncaglia, and D.A. Wisniacki, Phys. Rev. E \textbf{91}, 010902 (2015).

\bibitem{bib:rigol2008} M. Rigol, V. Dunjko, and M. Olshanii. Nature (London) \textbf{452}, 7189 (2008).

\bibitem{bib:santos2011} L.F. Santos, A. Polkovnikov, and M. Rigol, Phys. Rev. Lett. \textbf{107}, 040601 (2011).

\bibitem{bib:fazio2011} E. Canovi, D. Rossini, R. Fazio, G.E. Santoro, and A. Silva, Phys. Rev. B \textbf{83}, 094431 (2011).

\bibitem{bib:qpt2011} J. Simon, W.S. Bakr, R. Ma, M.E. Tai, P.M. Preiss, and M. Greiner, Nature (London) \textbf{472}, 307 (2011).

\bibitem{bib:tannor1993} J. Somloi, V. Kazakov, and D. Tannor, Chem. Phys. \textbf{172}, 85-98 (1993).

\bibitem{bib:rabitz1998} W. Zhu, J. Botina and H. Rabitz, J. Chem. Phys. {\bf 108}, 1953 (1998).

\bibitem{bib:krotov1996} V. F. Krotov, \textit{Global Methods in Optimal Control Theory} (Marcel Dekker, New York, 1996)

\bibitem{bib:sherson2013} M.C. Tichy, M.K. Pedersen, K. Molmer and J.F. Sherson, Phys. Rev. A {\bf 87}, 063422 (2013).

\bibitem{bib:calarco2015} I. Brouzos, A.I. Streltsov, A. Negretti, R.S. Said, T. Caneva, S. Montangero, and T. Calarco, Phys. Rev. A \textbf{92}, 062110 (2015).

\bibitem{bib:doria2011} P. Doria, T. Calarco, and S. Montangero, Phys. Rev. Lett. \textbf{106}, 190501 (2011).

\bibitem{bib:mintert2010} F. Platzer, F. Mintert, and A. Buchleitner, Phys. Rev. A \textbf{105}, 020501 (2010).

\bibitem{bib:caneva2012} T. Caneva, T. Calarco, and S. Montangero, New J. Phys. \textbf{14}, 093041 (2012).

\bibitem{bib:montangero2016} N. Rach, S. Montangero, and M. Paternostro, arxiv:165.07476 (2016).

\bibitem{bib:caneva2009} T. Caneva, M. Murphy, T. Calarco, R. Fazio, S. Montangero, V. Giovannetti, and G. E. Santoro, Phys. Rev. Let. {\bf 103}, 240501 (2009).

\bibitem{bib:murphy2010} M. Murphy, S. Montangero, V. Giovannetti, and T. Calarco, Phys. Rev. A \textbf{82}, 022318 (2010).

\bibitem{bib:hegerfeldt2013} G.C. Hegerfeldt, Phys. Rev. Lett. {\bf 111}, 260501 (2013).

\bibitem{bib:nos_qsl2013} P.M. Poggi, F.C. Lombardo and D.W. Wisniacki, Europhys. Lett. {\bf 104}, 40005 (2013).

\bibitem{bib:nos_qsl2015} P.M. Poggi, F.C. Lombardo and D.W. Wisniacki, J. Phys. A: Math. Theor. \textbf{48} 35FT02 (2015).

\bibitem{bib:sherson2016} J.J. Sørensen,	M.K. Pedersen,	M. Munch,	P. Haikka,	J.H. Jensen,	T. Planke,	M.G. Andreasen,	M. Gajdacz,	K. Mølmer,	A. Lieberoth, and J.F. Sherson, Nature (London) \textbf{532}, 210-213 (2016).

\bibitem{bib:caneva2014} T. Caneva, A. Silva, R. Fazio, S. Lloyd, T. Calarco, and S. Montangero, Phys. Rev. A \textbf{89}, 042322 (2014).

\bibitem{bib:lloyd2014} S. Lloyd, and S. Montangero, Phys. Rev. Lett. \textbf{113}, 010502 (2014).


\bibitem{bib:bethe} H.A. Bethe, Z. Phys. \textbf{71}, 205 (1931).

\bibitem{bib:vedral2001} M. C. Arnesen, S. Bose, and V. Vedral, Phys. Rev. Lett. \textbf{87}, 017901 (2001).

\bibitem{bib:fazio2002} A. Osterloh, L. Amico, G. Falci, and R. Fazio, Nature (London) \textbf{416}, 608-610 (2002).

\bibitem{bib:santos2012} L.F. Santos, F. Borgonovi, and F.M. Izrailev, Phys. Rev. E \textbf{85}, 036209 (2012).

\bibitem{bib:stockman} H.J. St\''ockmann, {\it Quantum chaos: an introduction}, (Cambridge University Press, Cambridge, 2006).

\bibitem{bib:bohigas}  O. Bohigas, in {\it Chaos and Quantum Physics Proceedings of the Les-Houches Summer School}. Session LII, ed. by M.J.Giannoni et.al (North Holland, Amsterdam, 1991).

\bibitem{bib:bose2007} S. Bose, Contemp. Phys. \textbf{48}, 13 (2007).

\bibitem{bib:spin1} X. Wang, A. Bayat, S.G. Schirmer, and S. Bose, Phys. Rev. A \textbf{81}, 032312 (2010). 

\bibitem{bib:spin2} D. Burgarth, K. Maruyama, M. Murphy, S. Montangero, T. Calarco, F. Nori, and M.B. Plenio, Phys. Rev. A \textbf{81}, 040303(R) (2010).

\bibitem{bib:spin2b} K. Korzekwa, P. Machnikowski, and P. Horodecki, Phys. Rev. A \textbf{89}, 062301 (2014).

\bibitem{bib:spin2c} S. Ashhab, Phys. Rev. A \textbf{92}, 062305 (2015).

\bibitem{bib:sklarz2002} S.E. Sklarz and D.J. Tannor, Phys. Rev. A \textbf{66}, 053619 (2002).

\bibitem{bib:montangero2007} S. Montangero, T. Calarco and R. Fazio, Phys. Rev. Lett. \textbf{99} 170501, (2007).

\bibitem{bib:werschnik2007} J. Werschnik, and E. Gross, J. Phys. B: At. Mol. Opt. Phys. \textbf{40} R175, (2007).

\bibitem{bib:nos_oc2015} P.M. Poggi, F.C. Lombardo and D.A. Wisniacki, Phys. Rev. A \textbf{92}, 053411 (2015).

\bibitem{bib:rabitz1998_2} W. Zhu and H. Rabitz, J. Chem. Phys. \textbf{109}, 385 (1998).

\bibitem{bib:tesch2002} C.M. Tesch and R. de Vivie-Riedle, Phys. Rev. Lett. \textbf{89}, 157901 (2002).

\bibitem{bib:ipr1} F. Evers and A.D. Mirlin, Phys. Rev. Lett. \textbf{84}, 3690 (2000).

\bibitem{bib:ipr2} E. Canovi, D. Rossini, R. Fazio, G.E. Santoro, and A. Silva, Phys. Rev. B \textbf{83}, 094431 (2011).

\bibitem{bib:brody} M.V. Berry and M. Robnik, J. Phys. A: Math. Gen. \textbf{17}, 2413 (1984).

}
\end{thebibliography}
\end{document}